\documentclass[%
 reprint,
superscriptaddress,
 amsmath,amssymb,
 aps,
floatfix,
]{revtex4-1}

\usepackage{dcolumn}
\usepackage{bm}

\usepackage{graphicx} 
\usepackage{color}

\usepackage{ulem} 

\usepackage[dvipdfmx,
colorlinks=true, citecolor=blue, urlcolor=blue, linkcolor=blue,
setpagesize=false, bookmarks=false
]{hyperref}

\begin{document}
\title{Controlling inversion and time-reversal symmetries by subcycle pulses in the one-dimensional extended Hubbard model}
\author{Kazuya Shinjo}
\affiliation{Computational Quantum Matter Research Team, RIKEN Center for Emergent Matter Science (CEMS), Wako, Saitama 351-0198, Japan}
\author{Shigetoshi Sota}
\affiliation{Computational Materials Science Research Team,
RIKEN Center for Computational Science (R-CCS), Kobe, Hyogo 650-0047, Japan}
\affiliation{Quantum Computational Science Research Team,
RIKEN Center for Quantum Computing (RQC), Wako, Saitama 351-0198, Japan}
\author{Seiji Yunoki}
\affiliation{Computational Quantum Matter Research Team, RIKEN Center for Emergent Matter Science (CEMS), Wako, Saitama 351-0198, Japan}
\affiliation{Computational Materials Science Research Team,
RIKEN Center for Computational Science (R-CCS), Kobe, Hyogo 650-0047, Japan}
\affiliation{Quantum Computational Science Research Team,
RIKEN Center for Quantum Computing (RQC), Wako, Saitama 351-0198, Japan}
\affiliation{Computational Condensed Matter Physics Laboratory,
RIKEN Cluster for Pioneering Research (CPR), Saitama 351-0198, Japan}
\author{Takami Tohyama}
\affiliation{Department of Applied Physics, Tokyo University of Science, Tokyo 125-8585, Japan}

\date{\today}
             

\begin{abstract}
Owning to their high controllability, laser pulses have contributed greatly to our understanding of strongly correlated electron systems.
However, typical multicycle pulses do not control the symmetry of systems, which plays an important role in the emergence of novel quantum phases.
Here, we demonstrate that subcycle pulses whose oscillation is less than one period within a pulse envelope can control inversion and time-reversal symmetries in the electronic states of the one-dimensional extended Hubbard model.
Using an ultrashort subcycle pulse, one can generate a steady electric current (SEC) in a photoexcited state due to an Aharonov-Bohm flux instantaneously introduced through the phase of an electric field. 
Consequently, time-reversal symmetry is broken. In contrast, a broad subcycle pulse does not induce SEC but instead generates electric polarization, thus breaking inversion symmetry. Both symmetry breakings in a photoexcited state can be monitored by second harmonic generation. 
These findings provide a new methodology for designing the symmetries of electronic states and open up a new field of subcycle-pulse engineering.
\end{abstract}
\maketitle


%
\section{Introduction}
Inversion- and time-reversal-symmetry breakings can induce novel properties such as magnetoelectric effects, which are significant for applications.
Multiferroic magnets~\cite{Wang2009, Tokura2014, Spaldin2019}, Weyl semimetals~\cite{Witczak-Krempa2014, Armitage2018}, and quantum liquid crystals~\cite{Kivelson1998, Vojta2009, Fradkin2010, Fradkin2012} that break these symmetries have attracted much attention.
On the other hand, there have been several attempts to control these symmetries with external electric fields: time-reversal symmetry can be broken by inducing an electric current in semiconductors, semimetals, and superconductors with dc and terahertz electric fields~\cite{Khurgin1995, Ruzicka2012, Cheng2014, Tokman2019, Vaswani2020, Takasan2021, Sirica2022}.
In addition, the development of pulse lasers has opened up a new paradigm in strongly correlated electron systems.
Ultrashort pulses of 6~fs comparable to a time scale of electron hopping induce a current in a superconductor usually hindered by thermalization due to electron scattering~\cite{Kawakami2020}.

With typically used multicycle pulses, inversion and time-reversal symmetries are difficult to control. 
Here, we address a question of whether high-field and ultrashort subcycle pulses having oscillations less than one period within a pulse envelope can change these symmetries.

To answer this question, we investigate inversion- and time-reversal-symmetry breakings in the one-dimensional (1D) extended Hubbard model (1DEHM) with a subcycle pulse being applied.
We numerically demonstrate that steady electric current (SEC) is induced, thus time-reversal symmetry is broken, in a photoexcited state when the ultrashort subcycle pulse introduces an instantaneous change in flux. 
This is because unlike a multicycle pulse, the subcycle pulse has the ability to generate Aharonov-Bohm flux~\cite{Aharonov1959}.
Since the SEC found is proportional to the Drude weight in the long-time limit, time-reversal symmetry can be broken not only in metals but also in photon-absorbed Mott insulators. 
Therefore, we can control time-reversal-symmetry breaking by tuning the flux generated through the phase of an electric field within a pulse envelope, i.e., carrier-envelope phase (CEP) $\phi_\text{CEP}$.
The symmetry breaking is evinced by the emergence of current-induced second-harmonic generation (SHG) and optical rectification (OR).
Furthermore, we find that a broad subcycle pulse with large intensity, i.e., a high-field terahertz pulse, not only induces quantum tunneling but also produces a polarization in a Mott insulator.
Therefore, inversion-symmetry breaking associated with the induced polarization and the resulting glassy dynamics~\cite{Shinjo2022} is maintained after pulse irradiation, leading to the emergence of polarization-induced SHG (PSHG) and OR.

The rest of this paper is organized as follows. 
In Sec.~\ref{sec2}, we introduce the 1DEHM and briefly explain the time-dependent density-matrix renormalization group (tDMRG) method.
We show the numerical results for the 1DEHM excited by a subcycle pulse in Sec.~\ref{sec3}.
In Sec.~\ref{sec3a}, we numerically demonstrate that SEC is induced by a ultrashort subcycle pulse.
The consequent time-reversal symmetry breaking is monitored by SHG in Sec.~\ref{sec:shg}.
In addition, we demonstrate in Sec.~\ref{sec:inversion} that the inversion symmetry is broken by a broad subcycle pulse, which is contrasted with the case for photoexcitations with a milticicle pule in Sec.~\ref{sec:multi}.
Finally, we give a summary in Sec.~\ref{sec4}.
The SEC induced by a flux quench, in a 1D superconducting state, and in a photo-absorbed Mott insulator is also discussed in the appendixes.

\section{Model and method}\label{sec2}
To investigate nonequilibrium properties of a 1D Mott insulator, we consider the 1DEHM with a vector potential $A(t)$ described by the following Hamiltonian: 
\begin{align}\label{Eq-Hamiltonian}
\mathcal{H}=&-t_\mathrm{h}\sum_{i,\sigma} B_{i,\sigma}
+ U\sum_i n_{i,\uparrow}n_{i,\downarrow} + V\sum_i n_{i}n_{i+1},
\end{align}
where $B_{i,\sigma}=e^{iA(t)} c_{i,\sigma}^\dag c_{i+1,\sigma}+\text{H.c.}$, $c_{i,\sigma}^{\dag}$ is the creation operator of an electron with spin $\sigma (= \uparrow, \downarrow)$ at site $i$, and $n_i=\sum_\sigma n_{i,\sigma}$ with $n_{i,\sigma}=c^\dagger_{i,\sigma}c_{i,\sigma}$.
Unless otherwise noted, we set $(U,V)=(10,3)$, taking the nearest-neighbor hopping $t_\mathrm{h}$ to be the unit of energy ($t_\mathrm{h}=1$), which is a typical value for describing 1D Mott materials such as ET-F$_{2}$TCNQ~\cite{Yamaguchi2021} and Sr$_{2}$CuO$_{3}$~\cite{Kishida2001}.
Spatially homogeneous electric field $E(t)=-\partial_{t}A(t)$ applied along a chain of $L$ sites is incorporated via the Peierls substitution in the hopping terms~\cite{Peierls1933}.
In this paper, we consider two kinds of subcycle pulses.
One is an ultrashort subcycle pulse and the other is a broad one.
The former and latter pulses lead to a phase twist in momentum space and a potential tilt in real space, respectively.

The current density is given as 
\begin{align}
j(t)=\frac{it_\text{h}}{L}\sum_{j,\sigma}\langle e^{iA(t)} c_{j,\sigma}^\dag c_{j+1,\sigma} - e^{-iA(t)} c_{j+1,\sigma}^\dag c_{j,\sigma} \rangle_t ,
\end{align} 
where $\langle\cdots\rangle_t=\langle\psi(t)|\cdots|\psi(t)\rangle$ and $|\psi(t)\rangle$ is the time-dependent wave function at time $t$.
We employ the tDMRG method implemented by the Legendre polynomial~\cite{Shinjo2021, Shinjo2021b, Shinjo2022} under open boundary conditions (OBCs) and keep at most $\chi=3000$ largest density-matrix eigenstates.
To reduce finite-size effects, a potential term $Vn_{i}$ at the edges of the system at $i=1$ and $L$ is also introduced to the 1DEHM in Eq.~(\ref{Eq-Hamiltonian}).
We set the light velocity $c$, the elementary charge $e$, the Dirac constant $\hbar$, and the lattice constant to 1.

\section{Results}\label{sec3}

\subsection{Current generation}\label{sec3a}

\begin{figure}[t]
  \centering
    \includegraphics[clip, width=20pc]{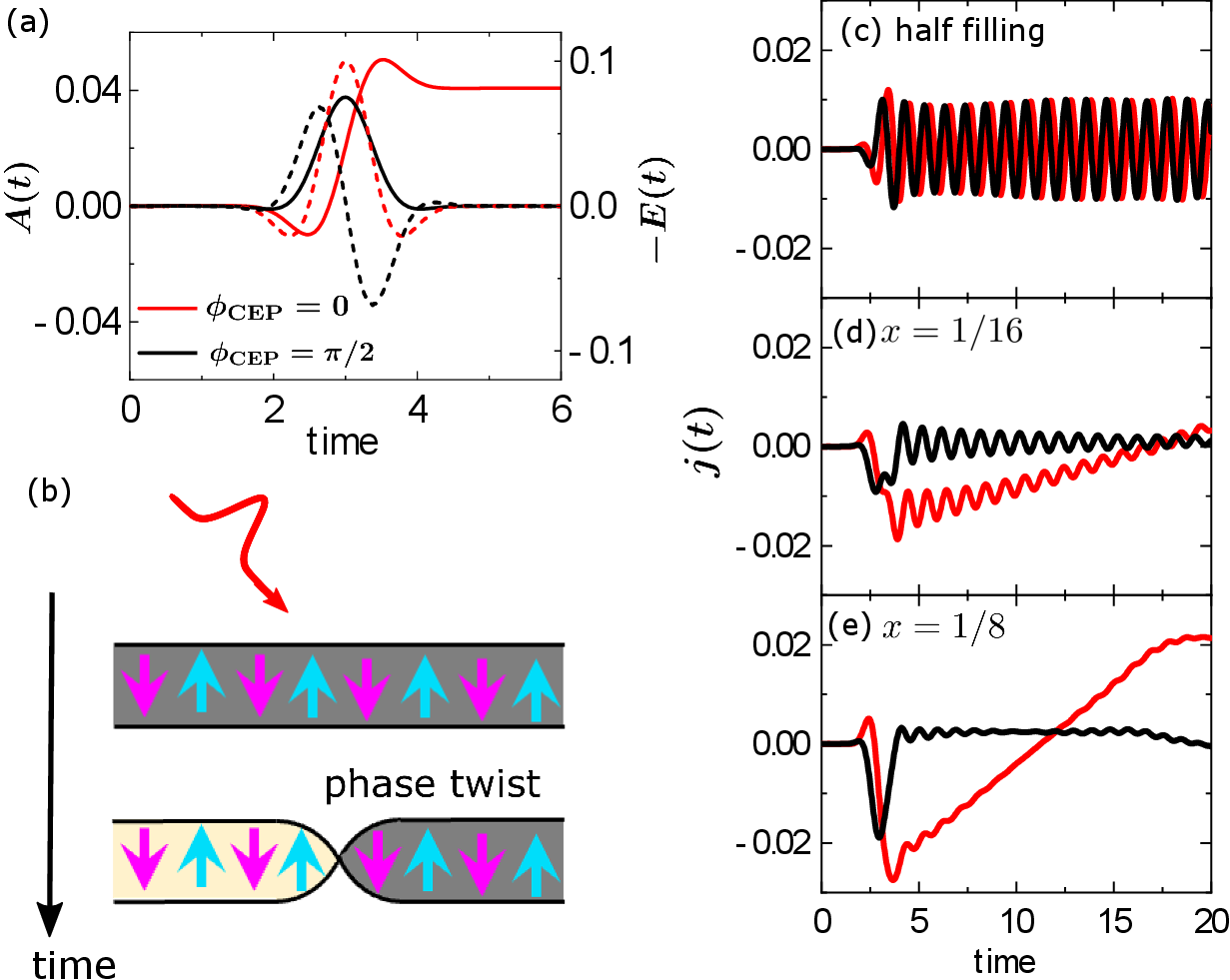}
    \caption{(a) $A(t)$ and $E(t)$ with $t_\text{d}=0.5$, $\Omega=3$, $t_{0}=3$, and $E_{0}=0.1$ shown by solid and broken lines, respectively. 
    Red and black lines are for $\phi_\text{CEP}=0$ and $\phi_\text{CEP}=\pi/2$, respectively. 
    (b) Schematic figure of instantaneous phase twist introduced by a ultrashort subcycle pulse. 
    Blue and purple arrows represent antiferromagnetically arranged spins. 
    (c)--(e) $j(t)$ for the 1DEHM with $(U,V)=(10,3)$ on $L=32$ at the concentration of electron doing $x=0$ (half filling), $1/16$, and $1/8$ evaluated after the ground state at $t=0$ is excited by the pulse shown in (a). 
    Red (black) lines denote the results for $\phi_\text{CEP}=0$ $(\pi/2)$.}
    \label{Fig1}
\end{figure}

The simplest system with electric current is a noninteracting metal with a dc electric field $E(t)= E_{0}$.
In this case, the induced current density exhibits Bloch oscillations~\cite{Raizen1997} characterized as $j(t)\propto \cos(E_\text{0}t)$. 
The problem consider here is rather opposite to this.
Namely, we apply an ultrashort pulse 
\begin{align}
E(t)=E_0 e^{-(t-t_0)^2/(2t_\mathrm{d}^2)} \cos\left[\Omega(t-t_0) + \phi_\text{CEP} \right]
\label{eq:e}
\end{align}
with small $t_\text{d}$. 
We can generate a subcycle pulse by tuning $t_\text{d}=0.5$ and $\Omega=3$ as shown in Fig.~\ref{Fig1}(a), where red and black solid (broken) lines indicate $A(t)$ [$E(t)$] for $\phi_\text{CEP}=0$ and $\pi/2$, respectively.
The time duration of the electric pulses is about 2. 
This corresponds to $2\times \hbar / t_\text{h} \approx 13$~fs and 4.3~fs for $t_\text{h}\approx 0.1$~eV in ET-F$_{2}$TCNQ~\cite{Wall2011} and $t_\text{h}\approx 0.3$~eV in Sr$_{2}$CuO$_{3}$~\cite{Kim2006, Schlappa2012}, respectively.
Tuning $\phi_\text{CEP}$ away from $\pi/2$, we can instantaneously twist the phase of wave functions by introducing $A(t)\neq 0$ after the pulse application, as schematically shown in Fig.~\ref{Fig1}(b).

Figures~\ref{Fig1}(c)--\ref{Fig1}(e) show the results of $j(t)$ evaluated after the ground state at $t=0$ is excited by a pulse shown in Fig.~\ref{Fig1}(a), for three different concentrations ($x$) of electron doping, i.e., $x=0$ (half filling), $1/16$, and $1/8$. 
At half filling, an oscillating current with a period $2\pi/\Delta$ is generated, where $\Delta \simeq6$ is the Mott gap for $(U,V)=(10,3)$. 
Similar oscillations remain even in finite $x$, although their amplitudes are small.
The amplitude is proportional to $E_{0}$ as far as $j(t)$ is small.
This oscillation is called Umklapp oscillation (UO) because of the following reason.
The Hamiltonian of the 1DEHM with a flux $A(t)=\phi$, denoted as $H_{\phi}$, is given from $H_{\phi=0}$ by a unitary transformation $H_{\phi} = U_{\phi} H_{\phi=0} U_{\phi}^{-1}$ with a twist operator $U_{\phi} = e^{i\phi P}$~\cite{Lieb1961}, where 
$P=\sum_{j}jn_{j}$ is a polarization operator. 
Since $U_{\phi}c_{j,\sigma}^{\dag}U_{\phi}^{-1}=e^{i\phi j}c_{j,\sigma}^{\dag}$ indicates a momentum shift $U_{\phi}\tilde{c}_{q,\sigma}^{\dag}U_{\phi}^{-1}=\tilde{c}_{q+\phi,\sigma}^{\dag}$ with $\tilde{c}_{q,\sigma}=\frac{1}{\sqrt{L}}\sum_{j=1}^{L}c_{j,\sigma}e^{-iqj}$, a flux quench induces an oscillation due to momentum shift picking up the effect of the Umklapp scattering~\cite{Nakagawa2016}. 

As shown in Fig.~\ref{Fig1}(c), we do not find a qualitative difference in $j(t)$ between $\phi_\text{CEP}=0$ and $\pi/2$ at half filling, except for a phase shift of the UO.
However, for electron-doped systems shown in Figs.~\ref{Fig1}(d) and \ref{Fig1}(e), we find distinct properties: the center of the UO is 0 for $\phi_\text{CEP}=\pi/2$, but not for $\phi_\text{CEP}=0$ leading to SEC.
At the time immediately after applying a pulse, $j(t)<0$ is kept for $\phi_\text{CEP}=0$.
After a certain time, the current reverses its direction leading to $j(t)>0$.
In OBCs, the flow direction of SEC alternates with a long periodicity $T\propto 1/L$, which forms a triangle-shaped slow oscillation in $j(t)$~\cite{Peotta2014}.
We note that only a quarter (half) of the triangle-shaped oscillation is shown by the red line in Fig.~\ref{Fig1}(d) [Fig.~\ref{Fig1}(e)].
This long periodicity corresponds to the fact that the Drude weight in Re$[\sigma(\omega)]$ appears at $\omega \neq 0$ under OBCs and its position approaches $\omega=0$ as $L$ increases~\cite{Rigol2008, Bellomia2020}.
The SEC can be characterized by, e.g., a root mean square $\left[\frac{1}{T}\int_{0}^{T}dtj(t)^{2}\right]^{1/2}$.

Figures~\ref{Fig1}(d) and \ref{Fig1}(e) indicate that an instantaneous phase twist making $A(t)\neq 0$ induces current when mobile carriers are already present in the initial ground state before the pulse irradiation.
Flux has been quenched in cold atoms with an artificial gauge field~\cite{Peotta2014} and with a sudden momentum shift~\cite{Mun2007}, but here we show that an ultrashort subcycle pulse with $t_\text{d}\lesssim1/t_\text{h}$ introduces a sudden change of flux in solids by tuning $\phi_\text{CEP}$.
We call this suddenly changed flux a flux pulse. 
Since an electric field of a flux pulse introduced by $A(t)=-\phi \theta(t)$ reads $E(t)=\phi \delta(t)$, i.e., $\tilde{E}(\omega)=\frac{1}{2\pi}\int_{-\infty}^{\infty}dt e^{i\omega t}E(t)=\phi/2\pi$, the linear response~\cite{Nakano1956, Kubo1957} of $j(t)$ to $\tilde{E}(\omega)$ is given by 
\begin{align}
j(t)\simeq \int_{-\infty}^{\infty} d\omega e^{-i\omega t} \tilde{E}(\omega)[\sigma_\text{sing}(\omega) + \sigma_\text{reg}(\omega)], 
\end{align}
where the optical conductivity $\sigma(\omega)$ is decomposed into the regular part $\sigma_\text{reg}(\omega)$ and singular part $\sigma_\text{sing}(\omega) =2\sigma_\text{D}i/(\omega+i0^{+})$ with the Drude weight $\sigma_\text{D}$.
In the long-time regime, where the regular part of conductivity is irrelevant, we obtain $j(t)\xrightarrow{t\to \infty} 2\sigma_\text{D}\phi$~\cite{Oshikawa2003, Peotta2014, Mierzejewski2014, Nakagawa2016}. 
This is a real-time representation of the Drude weight obtained by first taking the thermodynamic limit and then taking $t\rightarrow \infty$, which corresponds to the Drude weight obtained from the Kohn's theorem~\cite{Kohn1964, Shastry1990, Scalapino1992, Scalapino1993, Oshikawa2003}.
Note that this expression characterizes the SEC only when $j(t)$ is small, as in Figs.~\ref{Fig1}(d) and \ref{Fig1}(e), and thus nonlinear effects beyond the linear response regime are not dominant.
We have also confirmed that a flux pulse indeed plays the same role as a flux quench in inducing SEC (see Appendix~\ref{Aa}).

Notice also that the current expression obtained above is the same form as the London equation, although here we are dealing with a non-equilibrium state where $A(t)$ varies in time.
This implies that even at half-filling, it is possible to induce SEC by applying a flux pulse to a superconducting state with $U<0$, where a superfluid weight $\sigma_\text{S}\neq 0$. 
Since $\sigma_\text{D}$ and $\sigma_\text{S}$ are indistinguishable in 1D systems, we can still use the expression of the SEC obtained above by simply replacing $\sigma_\text{D}$ with $\sigma_\text{S}$. 
However, in this case, UO does not appear (see Appendix~\ref{Ab}).
Moreover, it is also possible to induce SEC in a Mott insulator if a photoexcited metallic state with $\sigma_\text{D}\neq 0$ is achieved before applying an ultrashort subcycle pulse (see Appendix~\ref{Ac}).

\subsection{Second-harmonic generation and optical rectification}\label{sec:shg}

\begin{figure}[t]
  \centering
    \includegraphics[clip, width=20pc]{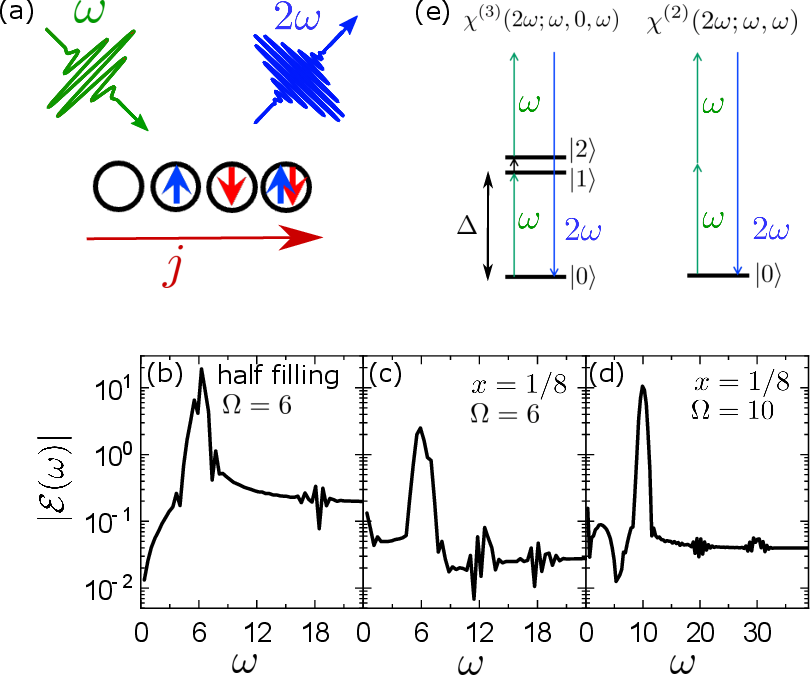}
    \caption{(a) Schematic figure of the CSHG in the 1DEHM. 
    (b)--(d) Harmonic generation $|\mathcal{E}(\omega)|$ for the 1DEHM with $(U,V)=(10,3)$ on $L=32$ at the concentration of electron doing (b) $x=0$ (half filling) and (c, d) $x=1/8$ after the initial ground state at $t=0$ is excited by a $\phi_{\rm CEP}=0$ flux pulse with $t_\text{d}=0.2$, $\Omega=0$, and $E_0=0.1$ applied at $t_{0}=1$. 
    The harmonic generation $|\mathcal{E}(\omega)|$ in (b, c) [(d)] is evaluated from the current density $j(t)$ in $2<t<18$ induced by a multicycle pulse with $t_\text{d}=2$, $E_0=1$, and $\Omega=6$ ($\Omega=10$) applied subsequently at $t_{0}=10$. 
    (e) Schematic diagrams representing $\chi^{(3)}(2\omega;\omega,0,\omega)$ and $\chi^{(2)}(2\omega;\omega,\omega)$ in the 1DEHM. The ground state $|0\rangle$ and two excitonic states $|1\rangle$ and $|2\rangle$ of the 1DEHM are indicated.}
    \label{Fig2}
\end{figure}

The induction of SEC breaks the time-reversal symmetry, which can change fundamental properties of electronic states such as the degeneracy of a Kramers pair~\cite{Kravtsov1992} and nonlinear optical responses~\cite{Boyd}.
Here, we detect the time-reversal symmetry breaking by SHG and OR, as schematically shown in Fig.~\ref{Fig2}(a).
In the 1DEHM, we numerically demonstrate that current-induced SHG (CSHG) actually emerges by applying a $\phi_\text{CEP}=0$ flux pulse at $t_{0}=1$ with $t_\text{d}=0.2$, $\Omega=0$, and $E_0=0.1$.
After completing the application of this subcycle pulse, we irradiate a multicycle pulse at $t_{0}=10$ with $t_\text{d}=2$ and $E_0=2$.
Figures~\ref{Fig2}(b)--\ref{Fig2}(d) show the resulting harmonic generation $|\mathcal{E}(\omega)|=|\omega \tilde{j}(\omega)|$ evaluated from $\tilde{j}(\omega)=\frac{1}{2\pi}\int dt e^{i\omega t}j(t)$.

Figures \ref{Fig2}(b) and \ref{Fig2}(c) show harmonic generation by $\Omega=6$ photons for half filling and $x=1/8$, respectively.
When the CSHG emerges, spectral weights should appear at $\omega= 2\Omega=12$ in the spectra.
This is indeed the case for $x=1/8$ in Fig.~\ref{Fig2}~(c), but not for half filling in Fig.~\ref{Fig2}(b).
These results are consistent with those shown in Figs.~\ref{Fig1}(c) and \ref{Fig1}(e), where the SEC is absent at half filling but present at $x=1/8$.
Figure~\ref{Fig2}(d) shows the same results as in Fig.~\ref{Fig2}(c), but a $\Omega=10$ pulse is used, resulting in a structure appearing at $\omega= 2\Omega=20$ due to the CSHG.
We have confirmed that the CSHG signals emerge also with other $\Omega$, but their intensities are smaller than that with $\Omega=6$.
This is understood because of a relation~\cite{Khurgin1995, Cheng2014, Ruzicka2012} $\chi^{(2)}(2\omega;\omega,\omega)\propto \chi^{(3)}(2\omega;\omega,0,\omega)j(t)$, where $\chi^{(l)}$ is the $l$-th order nonlinear optical susceptibility~\cite{Boyd}.
Schematic diagrams for these susceptibilities are shown in Fig.~\ref{Fig2}(e), assuming that the Mott-gap excitation remains for $x=1/8$. 
There are excitonic levels $|1\rangle$ and $|2\rangle$, which are transited from the ground state $|0\rangle$ of the 1DEHM with one- and two-photon processes, respectively.
Since the transition dipole moment $\langle1|x|2\rangle$ is anomalously large~\cite{Kishida2000, Kishida2001, Ono2004}, $\chi^{(3)}(2\omega;\omega,0,\omega)$ is enhanced by tuning $\omega$ to the gap energy $\Delta\simeq6$.
Moreover, we find spectral weights due to the OR at $\omega \sim 0$ for $x=1/8$ in Figs.~\ref{Fig2}(c) and \ref{Fig2}(d), but not for half filling in Fig.~\ref{Fig2}(b).

\subsection{Inversion-symmetry breaking}\label{sec:inversion}

\begin{figure}[t]
  \centering
    \includegraphics[clip, width=20pc]{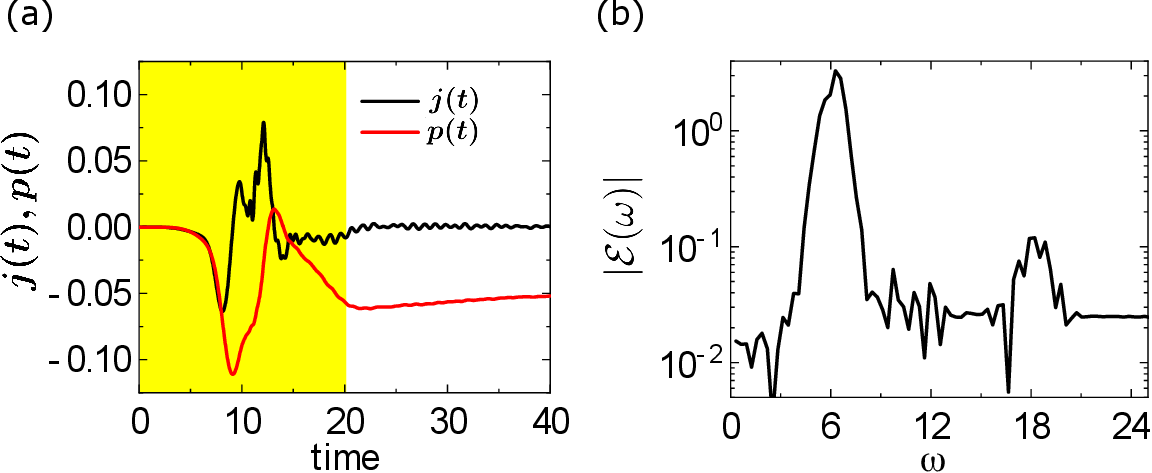}
    \caption{(a) $j(t)$ and $p(t)$, denoted by black and red lines, respectively, for the half-filled 1DEHM with $(U,V)=(10,3)$ on $L=32$ after the initial ground state at $t=0$ is excited by a monocycle $\Omega=0$ pulse with $t_\text{d}=2$ and $E_{0}=2$ applied at $t_{0}=10$. 
    The yellow shade is a guide to the eye to indicate the time during which the pulse is applied.
    (b) Harmonic generation $|\mathcal{E}(\omega)|$ for the photoexcited state prepared in (a) evaluated from $j(t)$ in $20<t<40$ that is induced by an $\Omega=6$ pulse with $t_\text{d}=2$ and $E_{0}=1$ applied subsequently at $t_{0}=30$.
    }
    \label{Fig3}
\end{figure}

As shown above, an ultrashort subcycle pulse cannot induce a photoexcited state with broken inversion- or time-reversal symmetry if the initial state is a Mott insulator.
However, changing the shape of a subcycle pulse to a broad and strong one (i.e., a pulse with larger $t_{\rm d}$ and $E_0$, respectively), we can induce inversion-symmetry breaking in a Mott insulator.
Figure~\ref{Fig3}(a) shows $j(t)$ and the polarization density $p(t)=\int_{0}^{t}ds j(s)$ for the half-filled 1DEHM on $L=32$.
Here, a broad monocycle $\Omega=0$ pulse with $t_\text{d}=2$ and $E_{0}=2$ is applied at $t_{0}=10$, which breaks down a Mott insulating state via quantum tunneling~\cite{Oka2003, Shinjo2022}.
Current is induced only during the pulse irradiation, indicated by a yellow shade in the figure, and is strongly suppressed afterward.
In contrast, polarization changes before and after the pulse irradiation and remains finite, $p(t)\neq 0$, even after the pulse disappears.
As discussed in Ref.~\cite{Shinjo2022}, the sum of doublon density and polarization is effectively conserved in a strong electric field.
After terminating the pulse irradiation, a doublon is essentially no longer generated (although the number of doublons still slightly fluctuats), thus leading to $j(t)=\partial_{t}p(t) \propto [\mathcal{H},P] \sim 0$.
The effective conservation of polarization causes glassy dynamics, and hence the inversion-symmetry breaking with $p(t)\simeq p_{0}\neq 0$ is not extinguished by thermalization for a certain time.

Consequently, as shown in Fig.~\ref{Fig3}(b), we find spectral weights at $\omega= 2\Omega=12$ in $|\mathcal{E}(\omega)|$. 
This is the PSHG, signaling inversion-symmetry breaking in a 1D Mott insulator.
In addition, we find spectral weights at $\omega \simeq 0$ due to the OR in Fig.~\ref{Fig3}(b).
These results suggest that it is possible to control the inversion symmetry by a terahertz pulse in Mott insulators, as has been done in ferroelectrics~\cite{Miyamoto2013, Miyamoto2018, Ohmura2019}.
We should note that spectral peaks of harmonic generations are broader in Fig.~\ref{Fig3}(b) than in Fig.~\ref{Fig2}(c).
This behavior may be due to a large number of photoinduced carriers in Fig.~\ref{Fig3}(b), which makes the optical conductivity of the photoexcited state significantly different from that of the ground state~\cite{Shinjo2022}.

We note that $A(t)\neq 0$ can be introduced after applying a broad subcycle pulse as well as an ultrashort subcycle pulse.
However, the effect of a phase twist becomes smaller as the width of a pulse becomes broader.
Given the broad limit of a subcycle pulse, an adiabatic introduction of $A(t)$ causes nothing physically since $A(t)$ is removed by a gauge transformation under OBCs.
Thus, the broad subcycle pulse hardly induce SEC in 1DEHM near half filling.
It is interesting to further investigate the effect of a phase twist due to a broad subcycle pulse as also examined in Ref.~\cite{Imai2022}, which is left for a future work.

\subsection{Multicycle-pulse excitation}\label{sec:multi}

\begin{figure}[t]
  \centering
    \includegraphics[clip, width=20pc]{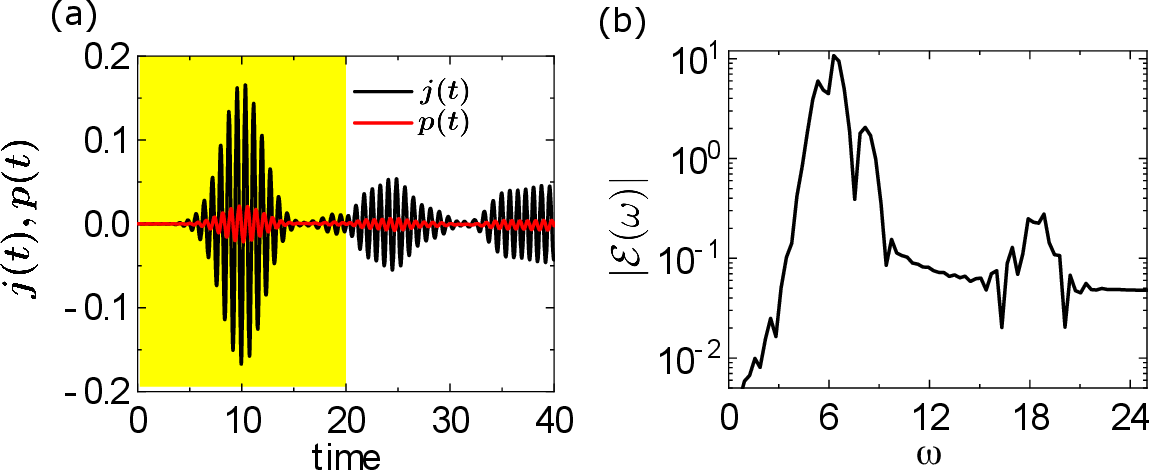}
    \caption{
    Same as Fig.~\ref{Fig3} but for a pump pulse with $\Omega=8$. 
    }
    \label{Fig_SHG_Omega8}
\end{figure}

In contrast to a subcycle pulse, a multicycle pulse cannot break inversion symmetry in a Mott insulator, leading to the absence of SHG and OR.
Figure~\ref{Fig_SHG_Omega8}(a) shows $j(t)$ and $p(t)$ for the half-filled 1DEHM excited by a multicycle $\Omega=8$ pulse with $t_\text{d}=2$ and $E_{0}=1$ applied at $t_0=10$.
This pulse with a photon energy larger than the Mott gap is absorbable and induces a metallic state. 
However, as shown in Fig.~\ref{Fig_SHG_Omega8}(b), we find no signal of SHG and OR at $\omega= 2\Omega=12$ and $\omega\sim0$ in harmonic generation $|\mathcal{E}(\omega)|$, respectively, where SHG and OR are evaluated from $j(t)$ in $20<t<40$ induced by an $\Omega=6$ pulse with $t_\text{d}=2$ and $E_{0}=2$ applied at $t_{0}=30$.
The absence of SHG and OR indicates no breaking of inversion and time-reversal symmetries. 
Note that the harmonic generation $|\mathcal{E}(\omega)|$ in both Fig.~\ref{Fig3}(b) and Fig.~\ref{Fig_SHG_Omega8}(b) is obtained for the photoexcited 1D Mott insulators with the same induced doublon density $\Delta n_\text{d}=0.1$, but there is significant difference in their symmetries. 
Here, the induced doublon density $\Delta n_\text{d}$ generated by a pump pulse is evaluated by 
$\Delta n_\text{d} := \frac{1}{L}\bigl[ \overline{\langle I \rangle_{t}} - \langle I\rangle_0 \bigr]$, where $I=\sum_{j}n_{j,\uparrow}n_{j,\downarrow}$ and $ \overline{\langle I \rangle_{t}}$ is the time average of $\langle I \rangle_{t} = \langle \psi(t)|I|\psi(t)\rangle$ taken from $t=21$ to $22$.

\section{Summary}\label{sec4}
We have investigated photoexcited states of the 1DEHM with a subcycle pulse. 
We have found that the SEC is induced in a photoexcited state of the 1DEHM by the flux pulse, i.e., an ultrashort subcycle pulse with a properly chosen CEP, instantaneously twisting the phase of a many-body wave function. 
Since the induced SEC is proportional to the Drude and superfluid weights, we can generate the SEC not only in metals, but also superconductors and photon-absorbed Mott insulators.
Although the SEC does not permanently flow due to thermalization in metallic states of nonintegrable systems, we expect that the SEC can still be observed because the ultrafast dynamics considered here is of a few femtoseconds before thermalization eventually occurs due to electron scattering. 
This implies that we can transiently control the time-reversal symmetry in photoexcited states by tuning the CEP of a flux pulse. 
The symmetry breaking can be detected by the emergence of CSHG. 

Furthermore, we have found that polarization is induced in a Mott insulator by a high-field terahertz pulse, i.e., a broad subcycle pulse, which induces quantum tunneling.
The induced polarization is the origin of glassy dynamics~\cite{Shinjo2022}, and thus is not extinguished by thermalization for a certain time.
This suggests that we can also transiently control the inversion symmetry in 1D Mott materials, which leads to the emergence of PSHG.
The proposed method of controlling the time-reversal and inversion symmetries in a photoexcited state by using a subcycle pulse provides a useful tool to investigate the effects of these symmetry changes on strongly correlated electron systems.


\begin{acknowledgments}
This work was supported by CREST (Grant No.~JPMJCR1661), the Japan Science and Technology Agency, by the Japan Society for the Promotion of Science, KAKENHI (Grants No.~17K14148, No.~18H01183, No.~19H01829, No.~19H05825, No.~21H03455, No.~21H04446, and No.~JP23K13066) from Ministry of Education, Culture, Sports, Science, and Technology (MEXT), Japan, and by JST PRESTO (Grant No.~JPMJPR2013). 
Numerical calculation was carried out using computational resources of HOKUSAI at RIKEN Advanced Institute for Computational Science, the supercomputer system at the information initiative center, Hokkaido University, the facilities of the Supercomputer Center at Institute for Solid State Physics, the University of Tokyo, and supercomputer Fugaku provided by the RIKEN Center for Computational Science through the HPCI System Research Project (Project IDs: hp170325 and hp220048).
\end{acknowledgments}

\appendix

\section{Steady electric current induced by a flux quench} \label{Aa}

\begin{figure}[t]
  \centering
    \includegraphics[clip, width=20pc]{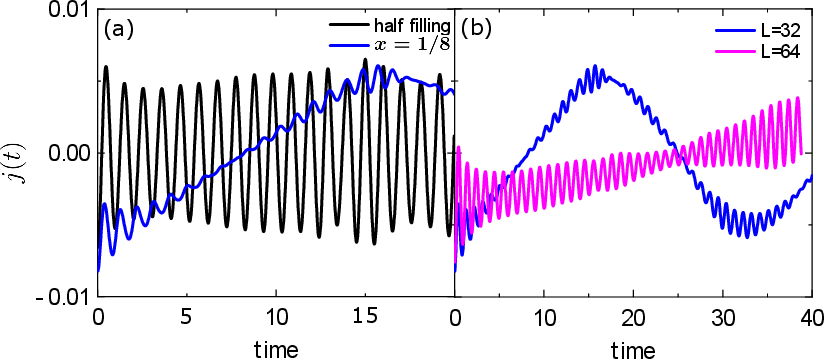}
    \caption{Electric current $j(t)$ for the 1DEHM with $(U,V)=(10,3)$ induced by a flux quench $A(t)=\phi\theta(t)$ with $\phi=0.01$ 
    applied to the initial ground state at $t<0$. 
    (a) $j(t)$ for $L=32$ at the concentration of electron doping $x=0$ (half filling) and $x=1/8$, indicated by the black and blue lines, 
    respectively. 
     (b) $L$ dependence of $j(t)$ at $x=1/8$. The blue line is the results for $L=32$, which are the same as those shown 
     by the blue line in (a), 
     and the magenta line is the results for $L=64$.}
    \label{FigS1}
\end{figure}

Figure~\ref{FigS1} shows the numerical results of $j(t)$ after quenching a flux $A(t)=\phi \theta(t)$ with $\phi=0.01$ for the initial ground state of the 1DEHM at $t<0$.
At half filling, $j(t)$ shows the UO with a period of $2\pi/\Delta$ around $j(t)=0$, as plotted by the black line in Fig.~\ref{FigS1}(a), where $\Delta\simeq6$ is the Mott gap for $(U,V)=(10,3)$.
On the other hand, for carrier density $x=1/8$ introduced by electron doping, the center of the UO deviates from $j(t)=0$, leading to the SEC [see the blue line in Fig.~\ref{FigS1}(a)].
Comparing these results in Fig.~\ref{FigS1}(a) with the red lines in Figs.~\ref{Fig1}(c) and \ref{Fig1}(e), we find that a flux pulse plays the same role as a flux quench in inducing the SEC.
We note that a flux pulse can be regarded as a flux quench only when the duration of the flux pulse $t_\text{d}$ is small compared with a timescale of the electron hopping $t_\text{h}$.
For large $t_\text{d}$, the generation of the SEC becomes unclear since the change of the flux in time is gradual. 
Furthermore, we show in Fig.~\ref{FigS1}(b) the $L$ dependence of the SEC induced by quenching a flux $\phi=0.01$ for the 1DEHM at $x=1/8$. 
The SEC has a long periodicity in time proportional to $1/L$ due to the OBCs.

\section{Steady electric current in a 1D superconducting state}\label{Ab}

\begin{figure}[t!]
  \centering
    \includegraphics[clip, width=20pc]{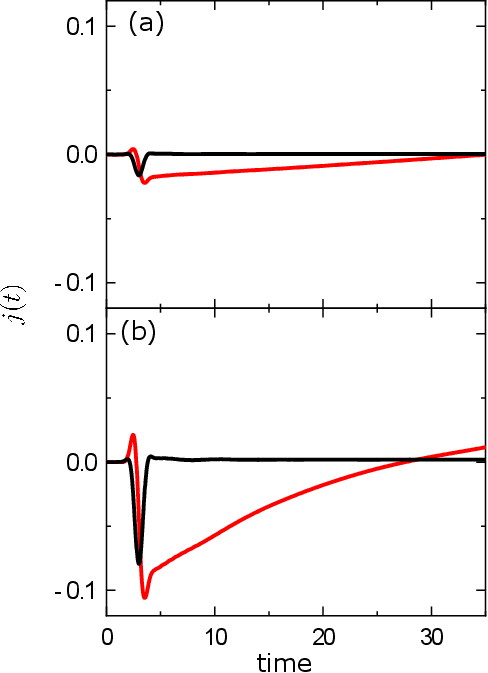}
    \caption{
    (a) Same as Fig.~\ref{Fig1}(c) but for the negative-$U$ Hubbard model with $U=-10$ at half filling.  
    (b) Same as (a) but with $E_{0}=0.5$.}
    \label{FigS2}
\end{figure}

Here, we demonstrate numerically that the SEC is induced in a superconducting state of the 1D half-filled negative-$U$ Hubbard 
model by applying a subcycle pulse. 
Figure~\ref{FigS2}(a) shows the results of $j(t)$ for the 1D half-filled Hubbard model with $U=-10$ after the ground state at $t=0$ is excited by the same subcycle pulse as in Figs.~\ref{Fig1}(c)--\ref{Fig1}(e).
Since there is a finite superfluid weight for the ground state of the 1D Hubbard model even at half filling, the SEC can be induced in the same way as in the 1DEHM with $(U,V)=(10,3)$.
The induced SEC becomes larger by applying a pulse with larger $E_{0}$, as shown in Fig.~\ref{FigS2}(b).
Notice also that the UO is absent in a photoexcited superconducting state, while it is present for the case of the 1DEHM with $(U,V)=(10,3)$.

\section{Steady electric current in a photon-absorbed Mott insulator}\label{Ac}

\begin{figure}[t!]
  \centering
    \includegraphics[clip, width=20pc]{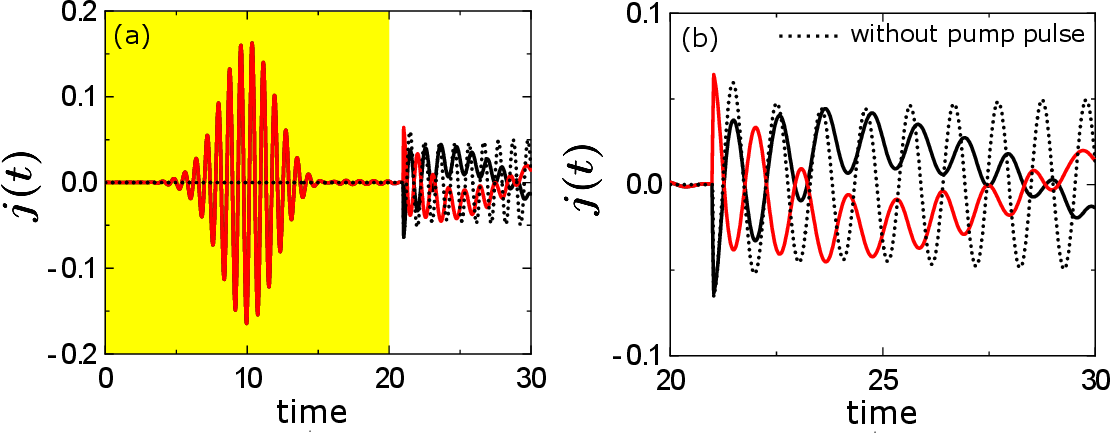}
    \caption{(a) $j(t)$ for the half-filled 1DEHM with $(U,V)=(10,3)$ on $L=48$ evaluated after the ground state at $t=0$ is 
    excited by a multicycle photon-absorbable pump pulse with $\Omega=8$, $E_{0}=1$, and $t_\text{d}=2$ at $t_{0}=10$. 
    The yellow shade is a guide for the eye to indicate the time during which the pump pulse is applied. 
    After metallization by the pump pulse, a flux pulse with $\Omega=0$, $E_0=0.1$, and $t_\text{d}=0.2$ is subsequently 
    applied at $t_{0}=21$. 
    The black and red solid lines are obtained with the pump and flux pulses for $\phi_\text{CEP}=0$ and $\pi$, respectively. 
    For comparison, $j(t)$ obtained without applying the pump pulse but only by applying the flux pulse with $\phi_\text{CEP}=0$ 
    is also shown by the black dotted line. 
    (b) Magnified view of (a) for $20<t<30$.}
    \label{Fig_photon_absorption}
\end{figure}

Here, we demonstrate numerically that it is possible to induce the SEC in a Mott insulator if a photoexcited metallic state with $\sigma_\text{D}\neq 0$ is achieved before applying an ultrashort subcycle pulse. 
For this purpose, we first apply a multicycle photon-absorbable pump pulse with $\Omega=8$, $E_{0}=1$, and $t_\text{d}=2$ at $t_{0}=10$ to the initial ground state of the half-filled 1DEHM with $(U,V)=(10,3)$ prepared at $t=0$, and then subsequently apply a $\phi_\text{CEP}=0$ (or $\pi$) flux pulse with $\Omega=0$, $E_{0}=0.1$, and $t_\text{d}=0.2$ at $t_{0}=21$.
Without the application of the pump pulse, the flux pulse alone cannot induce net current, but only the UO, as indicated by the dotted line in Fig.~\ref{Fig_photon_absorption}.

In contrast, when an insulator-to-metal transition is induced by the pump pulse, the center of the UO is shifted from 0 after applying the flux pulse, as shown by the black and red solid lines in Fig.~\ref{Fig_photon_absorption}.
The center of oscillating $j(t)$ for $\phi_\text{CEP}=0$ (black solid line) is positive for $22<t<27$ and that for $\phi_\text{CEP}=\pi$ (red solid line) is negative. This is because the applied flux has the opposite signs for $\phi_\text{CEP}=0$ and $\pi$.
$j(t)$ exhibits a more complicated structure than $j(t)$ obtained for the electron-doped systems, showing a triangle-like wave associated with current inversion [see Figs.~\ref{Fig1}(d) and \ref{Fig1}(e)], which is not clearly observed in Fig.~\ref{Fig_photon_absorption}(b).
This is due to the presence of low-energy spectral weights other than the Drude weight in the optical conductivity for the photon-absorbed Mott insulator of the half-filled 1DEHM excited by a near-infrared multicycle pulse~\cite{Lu2015, Rincon2021, Shinjo2022}.

\end{document}